\documentclass[12pt]{article}
\usepackage{graphics}
\usepackage{amssymb,epsfig,amsmath,euscript,array}
\usepackage{axodraw4j}
\usepackage{subfigure}
\usepackage{color}
\usepackage{cite}

\makeatletter
\@addtoreset{equation}{section}
\makeatother

\def\sst{\scriptscriptstyle}

\def\uno{\mbox{1 \kern-.59em {\rm l}}}

\let\vev=\Vev

\newcommand{\be}{\begin{equation}}
\newcommand{\ee}{\end{equation}}
\definecolor{holger}{rgb}{0,0.5,0.7}
\definecolor{edit}{rgb}{1,0,0}

\newcommand{\X}{{\sst X}}
\definecolor{durbeer}{rgb}{1,0,0}

\definecolor{durbeer2}{rgb}{0.8,0,0.5}

\newcommand{\Tr}{\text{Tr}}

\newcommand{\Nf}{N_{\mathrm{f}}}

\newcounter{multieqs}

\def\bd{\begin{document}}
\def\ed{\end{document}}
\def\nn{\nonumber}
\def\bea{\begin{eqnarray}}
\def\eea{\end{eqnarray}}
\let\bm=\bibitem
\let\la=\label

\begin{document}

\hfill{IPPP/09/49; DCPT/09/98}\\[-0.9cm]

\vspace{20pt}

\begin{center}

{\Large \bf Gaugino versus Sfermion Masses in Gauge Mediation}\\

\vspace{30pt}

{\bf  Steven A. Abel, Joerg Jaeckel and Valentin V. Khoze}

{\small \em {Institute for Particle Physics Phenomenology,
Durham University\\ Durham DH1 3LE, United Kingdom}}

\vspace{10pt}

{\sffamily \tt
s.a.abel@durham.ac.uk\\
joerg.jaeckel@durham.ac.uk
valya.khoze@durham.ac.uk}

\vspace{30pt}
\end{center}

\begin{abstract}
A well-known signature of supersymmetry breaking scenarios with ordinary
gauge mediation is a universal formula governing gaugino and sfermion masses
such that their ratio is of order one.
On the other hand, recently studied models with direct gauge mediation predict anomalously small ratios
of gaugino to scalar masses. It was argued that the smallness of gaugino masses is a consequence of being in the lowest energy state
of the SUSY-breaking low energy effective theory. To increase gaugino masses one either has to move to higher metastable vacuum or
alternatively remain in the original SUSY-breaking vacuum but extend the theory by introducing a lower-lying vacuum elsewhere.
We follow the latter strategy and show that the ratio of gaugino to sfermion masses can be continuously varied
between zero and of order one by bringing in a lower vacuum from infinity.
We argue that the stability of the vacuum is directly linked to the ratio between the gaugino masses and the underlying SUSY-breaking scale, i.e. the gravitino mass.

\end{abstract}

\setcounter{page}{0}
\thispagestyle{empty}
\newpage


Recently scenarios of gauge mediated supersymmetry
breaking~\cite{Dine:1981gu,Dine:1982zb,Nappi:1982hm,AlvarezGaume:1981wy}
have attracted renewed interest. Phenomenologically these models are
very predictive and at the same time have desirable features such as
automatic suppression of flavor changing neutral currents. Also with
an improved understanding of dynamical supersymmetry
breaking~\cite{ISS}, microscopic realisations have become more
accessible.

Gauge mediation scenarios are characterised by precise predictions for the mass spectrum of the gauginos and sfermions of the Standard Model.
There are two distinct gauge mediation scenarios discussed in the literature. The first one, ordinary gauge mediation with explicit messengers, predicts
a universal form for the gaugino and sfermion masses such that they are of the same
order (see, e.g.,~\cite{Giudice:1998bp}). A concrete realization of such mediation with explicit messengers
with the ISS model~\cite{ISS} as a SUSY breaking sector was given in \cite{Murayama:2006yf}.
In the alternative scenario of direct gauge mediation, the ratio of gaugino to sfermion masses was found to be small
in a wide class of models studied
in~\cite{Izawa:1997gs,Kitano:2006xg,Csaki:2006wi,Abel:2007jx,Abel:2007nr,Abel:2008gv}\footnote{In these models the leading order gaugino mass
was found to vanish. Taking into account subleading effects one typically finds $\lesssim 10^{-2}$ for the gaugino to
sfermion mass ratio~\cite{Abel:2007nr,Abel:2008gv}.}.
This corresponds to a (mildly) split SUSY.
In \cite{Komargodski:2009jf} it was argued that the smallness of the gaugino masses in all these models has a general origin. It is a consequence of expanding around
the lowest classical vacuum of the low energy effective theory.
Thus in order to avoid small gaugino masses one needs to be in an excited, metastable vacuum.
Inspection of ordinary gauge mediation (where gaugino masses are not small) shows that there one is indeed in such a metastable vacuum.
This demonstrates that metastability is directly related to the size of gaugino
masses.

The aim of this note is (a) to clarify and (to some degree) quantify this connection between gaugino masses and metastability, and (b)
to show that in dynamical gauge mediation models the ratio between gaugino and sfermion masses can be anywhere between zero and order one.
In other words within the gauge mediation setup one can continuously interpolate between the two extreme scenarios discussed above.
Below we construct and investigate various simple models to illustrate these points.

The strategy of extending the parameter space of ordinary gauge mediation has been pursued in the recent
literature~\cite{Shihextraordinary,Carpenter:2008wi,Meade:2008wd,Seiberg:2008qj,Buican:2008ws,Carpenter:2008he,Rajaraman:2009ga} largely in the context of
general gauge mediation~\cite{Meade:2008wd}. In particular, it was found that gaugino and sfermion masses are characterised by a priori different phenomenological
scales\footnote{General gauge mediation allows for a further split between the $SU(2)$, $SU(3)$ and $U(1)$ contributions to the gaugino and sfermion masses.
For simplicity, we will not concentrate on this feature.}.
In this paper we provide a weakly coupled dynamical implementation of this feature in the context of rather minimal ISS-based models. This also allows us to
directly connect phenomenological observables to the vacuum structure.

In the discussion above we have concentrated on the ratio of gaugino to sfermion masses. However, we will also argue that the most direct measure for the (meta-)stability
of the vacuum is actually the ratio between the gaugino mass and the underlying SUSY breaking scale which determines the gravitino mass.
Using this we can quantitatively relate the stability of the vacuum to the size of the measurable gaugino and gravitino masses.

Let us consider an O'Raifearthaigh model where SUSY is broken in the ground state. An example of such a model is provided by the perturbative magnetic
description of the ISS model.
If we expand around the ground state, gaugino masses are small, more precisely, the leading order contribution to gaugino masses vanishes~\cite{Komargodski:2009jf}.
To increase gaugino masses we can now follow two complementary strategies. We can either go to a higher metastable vacuum of this
theory as proposed by~\cite{Giveon:2009yu} (if the original theory did not have suitable excited vacua it needs to be deformed appropriately).
An alternative approach is to remain in the original ground state and extend the model in such a way as to bring in a lower lying vacuum from
infinity. This is a strategy we will follow to increase the gaugino mass but also to show that the gaugino mass can be increased \emph{continuously} by varying
parameters of the model. In the ISS context the latter strategy also allows us to make direct use of the established nice properties of original ISS vacuum
such as their cosmological viability~\cite{Abel:2006cr,Craig:2006kx,Fischler:2006xh,Abel:2006my}.

\noindent{\bf 1.} The first model we discuss is the original ISS construction with direct gauge mediation.

The model is an SQCD theory, with collider phenomenology
taking place in the magnetic description below a dynamical scale $\Lambda_{ISS}$.
There are $N_{f}$ flavours of quarks and
anti-quarks, $\varphi$ and $\tilde{\varphi}$ respectively, charged
under an $SU(N)_{\rm mg}$ magnetic gauge group, as well as an $N_{f}\times N_{f}$
meson $\Phi_{ij}$ which is a singlet under this gauge group.
The magnetic theory ìs weakly coupled in the IR and its electric dual is asymptotically free in the UV.

The ISS superpotential is given by
\begin{equation}
W_{\sst ISS}=h(\Phi_{ij}\varphi_{i}.\tilde{\varphi_{j}}-\mu_{ij}^{2}\Phi_{ji})\, .
\label{ISSsuperp}
\end{equation}
At the origin (in $\Phi$ space) $N$ of the magnetic quarks
get VEVs and the $SU(N)_{\rm mg}$ gauge symmetry is completely broken.
$N_f-N$ of the $F_\Phi$-terms remain non-zero however (thanks to the rank-condition)
and supersymmetry is broken with the vacuum being lifted at the origin
by (in the case of a degenerate and diagonal $\mu^2$ matrix) $\Delta V = (N_f-N)|\mu|^4$.
In addition to $W_{\sst ISS}$ there is a
non-perturbatively generated contribution to the superpotential \be W_{\rm dyn}\, =\, N\left(
\frac{\det_{\sst \Nf} h \Phi}{\Lambda_{\sst
ISS}^{\Nf-3N}}\right)^\frac{1}{N}\, , \label{Wdyn} \ee which introduces a
global supersymmetric minimum at large $\Phi$.

The flavour symmetry of the magnetic model is initially $SU(N_{f})$.
In order to do \emph{direct} gauge mediation, an $SU(5)_f$ subgroup of this symmetry
is gauged and identified with the parent $SU(5)$ of the Standard
Model, while an $SU(N)_{\rm mg}$ subgroup of $SU(N_f)$ is spontaneously broken by the
magnetic quark VEVs at the origin.
Note that by a gauge and
flavour rotation, the matrix $\mu_{ij}^{2}$ can be brought to a diagonal 2-5 form,
\begin{equation}
{\rm 2-5\,\, Model:} \qquad
\mu_{ij}^{2}=\left(\begin{array}{cc}
\mu_{Y}^{2}\mathbf{I}_{2} & 0\\
0 & \mu_{X}^{2}\mathbf{I}_{5}\end{array}\right) \, , \quad \mu_{Y}^{2} > \mu_{X}^{2}
\, ,
\label{25mu2}
\end{equation}
respecting the remaining $SU(2)_f\times SU(5)_f$ symmetries.

The spectrum in the magnetic description for the particular choice $N=2$ and $N_f=7$
is given
in Table~\ref{fieldstableM}.
Direct messengers include 2 $SU(5)$ (anti)fundamentals $\rho$ and $\tilde{\rho}$
(corresponding to the $SU(2)_{\rm mg}$ colour indices), along with 2 (anti)fundamentals $Z$ and $\tilde{Z}$
coming from the off-diagonal component of the original $7\times7$ mesons (corresponding to the $SU(2)_f$ flavour indices).
In addition the adjoint field $X$ can mediate as discussed in~\cite{Abel:2008gv}.

\begin{table}
\begin{center}
\begin{tabular}{|c|c|c|c|c|c|}
\hline
2-5 Model
&{\small $SU(2)_{\rm mg}$} &
{\small $SU(2)_f$}&
$SU(5)_f$& {\small $U(1)_{B}$} &
{\small $U(1)_{R}$}\tabularnewline
\hline
\hline
$\Phi_{ij}\equiv\left(\begin{array}{cc}
Y & Z\\
\tilde{Z} & X\end{array}\right)$&
{\bf 1}&
$\left(\begin{array}{cc}
Adj +{\bf 1} & \bar\square\\
\square & {\bf 1}\end{array}\right)$&
$\left(\begin{array}{cc}
{\bf 1} & \square\\
\bar{\square} & Adj+{\bf 1}\end{array}\right)$& 0 &
2
\tabularnewline
\hline
{\small $\varphi\equiv\left(\begin{array}{c}
\phi\\
\rho\end{array}\right)$}&
$\square$&
$\left(\begin{array}{c}
\bar{\square}\\ {\bf 1} \end{array}\right)$&
$\left(\begin{array}{c}
{\bf 1}\\
\bar{\square}\end{array}\right)$& $\frac{1}{2}$ &
$R$
\tabularnewline
\hline
{\small $\tilde{\varphi}\equiv\left(\begin{array}{c}
\tilde{\phi}\\
\tilde{\rho}\end{array}\right)$}&
$\bar\square$&
$\left(\begin{array}{c}
\square\\ {\bf 1} \end{array}\right)$&
$\left(\begin{array}{c}
{\bf 1}\\
\square\end{array}\right)$& $-\frac{1}{2}$ &
$-R$\tabularnewline
\hline
\end{tabular}
\end{center}
\caption{\it A 2-5 ISS Model. We show the ISS matter field decomposition under the gauge $SU(2)$, the flavour $SU(2)_f \times SU(5)_f$ symmetry,
and their charges under the $U(1)_B$ and $R$-symmetry. Both of the $U(1)$ factors above are defined as tree-level symmetries of the
magnetic ISS formulation in Eq.~\eqref{ISSsuperp}. The $R$-symmetry is anomalous. In the absence of deformations, the $R$-charges
of magnetic quarks, $\pm R$, are arbitrary. The baryon deformation fixes
$R=1$.
\label{fieldstableM}}
\end{table}

As noted in early work this model shows many promising features. First of course it breaks supersymmetry in
the metastable minimum at the origin, but supersymmetry is dynamically restored~\cite{ISS}. This
can be attributed to the fact that  the $R$-symmetry displayed in Table~\ref{fieldstableM} is anomalous, and
hence there exists a lower vacuum (or vacua) which is supersymmetric \cite{Nelson:1993nf}. At the same time, the supersymmetry breaking
minimum is cosmologically preferred~\cite{Abel:2006cr,Craig:2006kx,Fischler:2006xh,Abel:2006my} because one can argue that thermal effects would have driven the
early universe there.
Moreover, it was shown in Ref.~\cite{Abel:2008tx} that the Landau pole problem that usually plagues
direct gauge mediation can be avoided: this is because the ISS model itself runs into a Landau pole above which a
well-understood electric dual theory takes over. This results in a nett reduction in the effective number of messenger
flavours coupling to the SSM {\em above} the scale $\Lambda_{\sst ISS}$, and this in turn prevents the
Standard Model coupling running to strong coupling.

However, in the metastable vacuum the gaugino masses vanish (even in presence of \eqref{Wdyn}) because of an accidental $R$-symmetry, so we have
\be
 m_{\rm sc}\sim \frac{\alpha}{4\pi}  \frac{\mu^2_X}{\mu_{Y}}\,\,\, ; \, \, m_\lambda=0 \, .
\ee

\noindent{\bf 2.} Now we extend the minimal model above in order to achieve spontaneous $R$-symmetry breaking and consequentially non-vanishing gaugino masses.
$R$-symmetry breaking can be achieved radiatively by introducing appropriate deformations such as baryon deformations studied in~\cite{Abel:2007nr,Abel:2007jx}
or deformations involving
mesons~\cite{Abel:2008gv,Cho:2007yn}. Spontaneous $R$-symmetry breaking can also occur at tree-level~\cite{Carpenter:2008wi,Sun:2008va}.

For concreteness we briefly outline a simple baryon-deformed model,
\begin{equation}
W=\Phi_{ij}\varphi_{i}.\tilde{\varphi_{j}}-\mu_{ij}^{2}\Phi_{ji}+m_{\sst baryon}\varepsilon_{ab}
\varepsilon_{rs}\varphi_{r}^{a}\varphi_{s}^{b}\,\,
\label{Wbardef}
\end{equation}
where $i,j=1...7$ are flavour indices, $r,s=1,2$ run over the first
two flavours only, and $a,b$ are $SU(2)_{\rm mg}$ indices.
This is the superpotential of ISS with the
exception of the last term which is a baryon of the $SU(2)_{\rm mg}$ gauge group.

At tree-level this baryon-deformed model
has  a runaway to {\em broken} supersymmetry.
Parameterising the VEVs by
\bea
\label{phivevs}
\vev{\tilde\phi} &=& \xi\,\mathbf{I}_{2},\quad\quad\,\,\vev{\phi}=\kappa\,\mathbf{I}_{2},
\quad\quad\quad \vev{Y} = \eta\,\mathbf{I}_{2},\quad\quad\vev{X}=\chi\,\mathbf{I}_{5},
\eea
one finds~\cite{Abel:2007jx,Abel:2007nr} that the runaway is along the direction $\kappa = \mu_Y^2 /\xi$ and $\xi \rightarrow 0$.
Upon adding the Coleman-Weinberg
contributions  to the potential, the runaway direction is stabilized at large $\xi$ values where the
$R$-symmetry is spontaneously broken by a $\chi$ VEV that is of order $\mu_X$.

As before, the {\it direct} gauge mediation is implemented by
gauging the $SU(5)_{f}$ factor and identifying it
with the parent $SU(5)$ gauge group of the Standard Model.
The representations are given in Table~\ref{fieldstableM} but now $R=1$ is enforced by the baryon deformation.

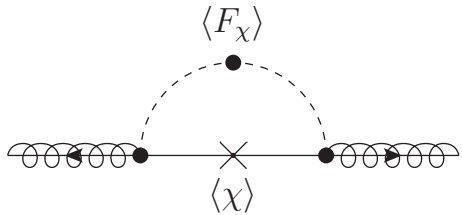
\begin{figure}[t]
\begin{center}
\begin{picture}(200,120)
\SetOffset(0,20)
\Gluon(0,0)(50,0){4.5}{5}
\ArrowLine(50,0)(0,0)
\Line(50,0)(120,0)
\Gluon(120,0)(170,0){4.5}{5}
\Vertex(50,0){3}
\Vertex(120,0){3}
\DashCArc(85,0)(35,0,180){3}
\ArrowLine(120,0)(170,0)
\Vertex(85,35){3}
\Text(85,50)[c]{\scalebox{1.1}[1.1]{$\langle F_{\chi}\rangle$}}
\Vertex(85,0){1}
\Line(80,5)(90,-5)
\Line(90,5)(80,-5)
\Text(85,-15)[c]{\scalebox{1.1}[1.1]{$\langle \chi\rangle$}}
\end{picture}
\end{center}
\vspace*{-0.5cm} \caption{\it One-loop contribution to the
gaugino masses. The dashed (solid) line is a bosonic (fermionic) messenger.
The blob on the scalar line indicates an insertion of $\langle F_{\chi}\rangle$ into the propagator
of the scalar messengers and the cross denotes an
insertion of the $R$-symmetry breaking VEV into the propagator of the fermionic messengers.}
\label{gauginofig}
\end{figure}

It was noted in Refs.~\cite{Abel:2007jx,Abel:2007nr,Abel:2008gv}
that the typical signature of this class of models is "split", with
 a hierarchical structure of SUSY breaking in the Standard Model sector
 of the form\footnote{Obtained from a complete one-loop~\cite{Abel:2007jx,Abel:2007nr} plus dominant higher loop contributions~\cite{Abel:2008gv}.}
\be
 m_{\rm sc}\sim \frac{\alpha}{4\pi}  \mu\,\,\, ; \, \, m_\lambda\sim10^{-2} m_{\rm sc} \, .
\ee
The reason for this hierarchy can be seen by looking at the leading order contributions to the
gaugino masses: they come from the one-loop diagram  shown in Fig.~\ref{gauginofig},
which has a \emph{single} insertion of an $R$-symmetry breaking VEV, and a SUSY breaking $F$-term, both of which
are required. However, because the functional form of the
leading order contribution to the gaugino mass
is proportional to a loop factor times $\partial V_{\rm tree}/\partial Y^{\star} = 0$, this contribution vanishes (see Sect. 3.2.1 of \cite{Abel:2008gv}).
The first non-vanishing contribution to gaugino masses therefore comes from
the one loop contribution to the potential perturbing this tree-level relation, or from
diagrams with at least 3 $F_\chi$ insertions, or alternatively from the mediating effects of the
pseudo-Goldstone $X$ modes. The suppression corresponds to typically an extra loop factor.
It is important to stress that the suppression of gaugino masses isn't directly attributable to the size of $R$-symmetry breaking --- indeed the $R$-symmetry
breaking VEVs are $\chi \sim \mu_X$ and not small.

\noindent{\bf 3.} In fact the authors of \cite{Komargodski:2009jf} gave a general argument which ties the smallness of the gaugino masses to the vacuum structure.
The models above are subject to this argument which we will briefly paraphrase.

For concreteness let us assume that the supersymmetry breaking sector is described by a renormalizable O'Raifeartaigh model (i.e. a general renormalizable Wess-Zumino
model which gets a non-vanishing F-term at the classical level).
One can then show~\cite{Ray:2006wk} that the superpotential can always be written
in the form\footnote{The transformation leading to this form of the superpotential is a linear change of field variables. The linearity
implies that the model remains renormalizable after and the Kaehler potential remains canonical. However, the field redefinition
obscures all symmetries that are broken along the $X$ direction.}
\begin{equation}
W=\zeta X+(\mu_{\alpha\beta}+\lambda_{\alpha\beta}X)\varphi_{\alpha}\varphi_{\beta}+\kappa_{\alpha\beta\gamma}\varphi_{\alpha}\varphi_{\beta}\varphi_{\gamma}.
\end{equation}
Here, $X$ is the goldstino superfield whose scalar component corresponds to a pseudo-Goldstone mode. This is a flat direction at tree-level.
The field $\varphi_{\alpha}$ contains both: fields that behave as messengers and others that are singlets under the standard model gauge group.
We refer to the messenger fields as $\varphi_{a}$ with a roman index.

Now in general the mass matrices for the bosonic and fermionic components of the messengers are given by
\begin{eqnarray}
m_{0,\,ac}^{2} & = & \left(\begin{array}{cc}
W^{a\beta}W_{\beta c} & W^{a\beta c}W_{\beta}\\
W_{a\beta c}W^{\beta } & W_{a\beta}W^{\beta c}\end{array}\right)\quad\quad\quad\quad
m_{1/2,\,ac}  =  W_{ac}
\end{eqnarray}

If we have a true vacuum the field $X$ must be stabilized by quantum corrections. Without loss of generality we can assume this to happen at $X=0$.
Now, we go along the classically flat $X$-direction where all the VEVs of the $\phi_{\alpha}$ are zero.
At leading order in $F_{\X}$ the contribution to the gaugino mass matrix is of the
form (see Fig.~\ref{gauginofig}),
\begin{eqnarray}\label{gauginomass}
\bar{m}_{\lambda}(X) & = &Tr((m_{1/2,\,ab})^{-1}W^{bcX}W_{X})\\\nonumber
& = &Tr((W_{ab})^{-1}W^{bcX}W_{X})
  =  Tr((\mu_{ab}+\lambda_{ab}X)^{-1}\lambda^{bc}\zeta),
\end{eqnarray}
where the physical gaugino masses $m_{\lambda}$ are obtained by including a factor $g^2/(16\pi^2)$ and an appropriate group theoretical factor.

Now consider the mass squared matrix of the scalar messengers.
\begin{eqnarray}
\nonumber
m_{0}^{2} & = &
\left(\begin{array}{cc}
W^{ab}W_{bc} & W^{ac\X}W_{\X}\\
W_{ac\X}W^{\X} & W_{ab}W^{bc}\end{array}\right)
\\\nonumber
 & = & \left(\begin{array}{cc}
W^{ab} & 0\\
0 & W_{ab}\end{array}\right)\left(\begin{array}{cc}
W_{bc} & (W_{bd})^{-1}W^{dc\X}W_{\X}\\
(W^{bd})^{-1}W_{dc\X}W^{\X} & W^{bc}\end{array}\right)\\
 & = & \left(\begin{array}{cc}
m_{1/2} & 0\\
0 & m_{1/2}\end{array}\right)\left(\begin{array}{cc}
m_{1/2} & \bar{m}_{\lambda}\\
\bar{m}_{\lambda} & m_{1/2}\end{array}\right)
\label{scalarmasses}
\end{eqnarray}
where at the last step we assume that all entries are real, and define
the matrix
\begin{eqnarray}
(\bar{m}_{\lambda})_{a}^{c} & = & (W_{ab})^{-1}W^{bc\X}W_{\X}
\end{eqnarray}
We also make the eigenvalues of $m_{1/2}$ all positive. Now we are
interested in what happens as we vary $X$ given that at some point
$\bar{m}_{\lambda}(X)\neq0$. The functional form of $\bar{m}_{\lambda}(X)$
is
\be
(\bar{m}_{\lambda})_{a}^{c}=(\lambda_{ab}X+\mu_{ab})^{-1}\lambda_{bc}\zeta
\ee
while the functional form of $m_{1/2}$ is
\be
(m_{1/2})_{ab}=(\lambda_{ab}X+\mu_{ab}).
\ee

If the leading order gaugino masses are non-vanishing in the original vacuum then $\bar{m}_{\lambda}(0)\sim(\mu^{-1}\lambda)\neq 0$.
Then since $\lambda^{ab}X+\mu_{ab}$ is linear in $X$
there is at least one root where an
$m_{1/2}$ eigenvalue vanishes.
Then in the scalar messenger mass squared matrix \eqref{scalarmasses} one can
perform a basis change such that it separates out a $2\times2$ block giving one positive and one negative eigenvalue (see also \cite{Komargodski:2009jf}).
The strategy is to first diagonalize $m_{1/2}$. Then we act on the scalar matrix by a basis change,
\begin{equation}
\left(
  \begin{array}{cc}
    A & 0 \\
    0 & B \\
  \end{array}
\right)\left(
         \begin{array}{cc}
           m^{2}_{1/2} & m_{1/2}m_{0} \\
           (m_{1/2}m_{0})^{\dagger} & m^{2}_{1/2} \\
         \end{array}
       \right)\left(
                \begin{array}{cc}
                  A^{-1} & 0 \\
                  0 & B^{-1} \\
                \end{array}
              \right)
              =\left(
                 \begin{array}{cc}
                     \begin{array}{cc}
                       0 & 0 \\
                       0 & \ddots \\
                     \end{array}
                    &
                        \begin{array}{cc}
                          x & 0 \\
                          0 & \ddots \\
                        \end{array}
                     \\
                        \begin{array}{cc}
                          x & 0 \\
                          0 & \ddots \\
                        \end{array}
                    &
                     \begin{array}{cc}
                       0 & 0 \\
                       0 & \ddots \\
                     \end{array}
                   \\
                 \end{array}
               \right)
\end{equation}
where the dots indicate blocks of non-vanishing elements. Hence,
we have a tachyonic messenger, implying the existence of  a lower lying vacuum where the
tachyonic scalar has a VEV\footnote{Strictly speaking the statement is that there
are tachyonic messengers: in the lower lying vacuum they may end up with a zero VEV. One way
for this to happen is if along the $X$ direction other particles become tachyonic first, as for example happens in the
model of \cite{Giveon:2009yu}.}.

The arguments above are clearly of a classical nature and one might wonder whether
quantum corrections could raise the erstwhile lower vacuum above the original one.
If this could be realised one could have a situation where the gaugino masses
are non-zero yet the vacuum is still the lowest lying non-supersymmetric one.
However this is unlikely. First of all, if the lower vacuum with tachyon condensation
were supersymmetric, then quantum corrections would be unable to lift it (since they
vanish where supersymmetry is restored). Hence this could only occur if the
lower lying vacuum were nonsupersymmetric. In this case  the Coleman-Weinberg contributions
along the classically flat $X$ direction can grow only logarithmically
with $X$:
\be
V_{CW}\sim \frac{1}{64\pi^2}\zeta^2 \log X\, ,
\ee
where $\sqrt{\zeta}$ is the typical SUSY breaking scale in the original vacuum.
Since one expects a generic lowering, $\Delta V_{\rm cl}$, of the vacuum to be of order $\zeta^2$,
one would have to have either very large $X$ (beyond the Planck scale) or
perform a fine-tuning of parameters to reduce $\Delta V_{cl}$ and allow $V_{CW}$ to dominate.

Another logical possibility for having sizable gaugino masses in the lowest vacuum is if the higher order contributions
in $F$ are relatively large. In the classes of simple models studied so far this does not happen.

\noindent{\bf 4.} Indirect (i.e. ordinary) gauge mediation gives a
familiar spectrum in which the gaugino masses are not suppressed and
are of the same order as the scalar ones. These theories too must
adhere to the above argument; that is they include lower lower lying
vacua along which the messenger fields get a VEV if they are to have
gaugino masses comparable to those of the scalars. A simple
realization \cite{Murayama:2006yf} of ordinary gauge mediation is to
couple explicit messengers $f$ and $\tilde{f}$ to the meson in the
magnetic ISS model \eqref{ISSsuperp}:
\begin{equation}
W_{\sst OGM}=h(\Phi_{ij}\varphi_{i}.\tilde{\varphi_{j}}-\mu_{ij}^{2}\Phi_{ji}) + \lambda'_{ij}  \Phi_{ij} \tilde{f} f+M_{f}\tilde{f}f \, .
\label{OGM}
\end{equation}
Clearly, all terms in this magnetic superpotential are renormalizable. From the point of view of the electric theory formulation the interaction
of messengers with mesons is a dimension 4 operator and thus $\lambda^{\prime}\sim \Lambda_{ISS}/M$ where $M$ is a high scale of new physics.
For $\lambda^{\prime},M_{f}\neq 0$ $R$-symmetry is explicitly broken.

At $\Phi=0$ there is a metastable SUSY breaking vacuum. In this vacuum, $F\sim\mu^2_{X}$ and the messenger mass is $M_{f}$.
Accordingly we find non-vanishing gaugino and scalar masses with the expected signature
of ordinary gauge mediation,
\begin{equation}
m_{\lambda}\sim \frac{\alpha}{4\pi}\frac{\lambda^{\prime}\mu^2_{X}}{M_{f}}\sim m_{sc}.
\end{equation}

Inspection shows that going to a suitably large VEV for $\Phi$ in the $X_{ij}$ direction (cf. Table~\ref{fieldstableM})
$X\sim M_{f}/\lambda^{\prime}$ the messengers $f$ become
tachyonic as expected from the general argument reviewed earlier. For the special case $(\mu^2_{X})_{ij}\sim \mu^2_{X}\delta_{ij}$ this lower
lying vacuum is actually a supersymmetric one as one would have expected from the broken $R$-symmetry. However, it should be noted
that this model has non-generic superpotential (e.g. in \eqref{OGM} there are no terms $\sim\Phi^2$) and the Nelson-Seiberg theorem does not apply.
For example for non-degenerate $\mu^{2}_{X}$ this lower vacuum is also non-supersymmetric.
This non-genericity of the potential is completely natural when starting from the electric formulation of the theory as
pointed out in~\cite{Murayama:2006yf} where the coefficient of the $\Phi^2$ terms is small and their effect is negligible in the vicinity of the SUSY breaking vacuum.

To summarize, ordinary gauge mediation has exactly the vacuum structure commensurate with its non-vanishing leading order gaugino masses.
Further note that conversely one can send the lower lying vacuum to infinity by taking $\lambda^{\prime}\to 0$. The price to pay is that the gaugino mass goes to
zero at the same time, $X\sim 1/(m_{\lambda})$. However, the ratio between gaugino and sfermion masses remains of order 1.
This shows that the fundamental connection is between the vacuum structure and the gaugino mass in relation to the underlying SUSY breaking scale
and not directly to the ratio between gaugino and sfermion masses. We will return to this point later.

\noindent{\bf 5.} Now we are ready to construct a model where changing the vacuum structure in this manner \emph{does} affect the ratio of gaugino to sfermion masses.
From the models discussed in {\bf \S1} and {\bf \S4} one can see that a hybrid model would provide for an interpolation.
This hybrid model gives a simple example where the ratio between gaugino and sfermion masses can be varied continuously as the distance in field space
to the lower vacuum is varied.

Specifically the superpotential is very similar to the ordinary gauge mediation:
\begin{equation}
W_{\sst hybrid}=h(\Phi_{ij}\varphi_{i}.\tilde{\varphi_{j}}-\mu_{ij}^{2}\Phi_{ji}) + \lambda'  \Tr(\Phi) \tilde{f} f+M_{f}\tilde{f}f \, .
\label{OGM}
\end{equation}
However, the messenger sector of this model is larger because it contains the direct messengers of {\bf \S1}
(in the notation of Table~\ref{fieldstableM}) $\rho,\tilde{\rho},Z,\tilde{Z}$
as well as the explicit messengers $f,\tilde{f}$ of {\bf \S4}.
Even though we call this a hybrid model it is essentially just a direct mediation model where we simply gauged a global symmetry
of all the matter fields including $f,\tilde{f}$.
This model could be easily extended by a variety of baryon and meson deformations which, however, will not give qualitative changes.

Now let us turn to gaugino and scalar masses. Both of them are well approximated by the leading order expressions in $F$.
Leading order gaugino masses receive contributions only from $f\tilde{f}$ messengers whereas scalar masses receive contributions from all messengers,
\begin{eqnarray}
m_{\lambda}\!\!&\sim&\!\!N_{\sst fm}\frac{\alpha}{4\pi}\frac{\lambda^{\prime}\mu^2_{X}}{M_{f}}\\\nonumber
m^{2}_{sc}\!\!&\sim&\!\!\left(\frac{\alpha}{4\pi}\right)^2\left[N_{\sst fm}\left(\lambda^{\prime}\frac{\mu^2_{X}}{M_{f}}\right)^2
+N_{\sst dm}\left(\frac{\mu^2_{X}}{\mu_{Y}}\right)^2\right],
\end{eqnarray}
where we have allowed for a number $N_{\sst fm}$ of explicit messengers and $N_{\sst dm}$ counts the effective number of direct messengers.

The ratio we are after is
\begin{equation}
\frac{m_{\lambda}}{m_{sc}}\sim
\frac{N_{\sst fm}\lambda^{\prime}}{\left[N_{\sst fm}\lambda^{\prime\,2}
+N_{\sst dm}\left(\frac{M^{2}_{f}}{\mu^2_{Y}}\right)\right]^{1/2}}.
\end{equation}
At $\lambda^{\prime}=0$ the ISS vacuum is stable\footnote{We ignore the non-perturbative superpotential.} and the supersymmetry is infinitely split.
At non-zero $\lambda^{\prime}$ a lower lying vacuum appears and moves closer as $\lambda^{\prime}$ increases. At the same time the ratio
also increases until it reaches values $\sim 1$.

Finally we note that the hybrid model as well as the ordinary gauge mediation model discussed in {\bf \S 4} are cosmologically
viable around the ISS-like vacuum where we break SUSY. This is because in the lower lying vacuum mesons as well as messengers develop VEVs.
Thus even the Standard Model gauge bosons are heavy there. In the early hot phase this excess of heavy degrees of freedom drives the universe
away from this vacuum towards the ISS-like vacuum near the origin~\cite{Abel:2006cr,Abel:2006my}.

\noindent{\bf 6.} Earlier in {\bf \S3} we have seen that there is a connection between gaugino masses and the existence of a lower vacuum.
Now we can go one step further and quantitatively relate the gaugino mass to the distance to this lower vacuum.
We note that in all the arguments above only the gaugino mass and the size of the $F$ terms played a role. Therefore, the quantitative
connection between observable parameters and the vacuum stability will involve the gaugino mass and the gravitino mass (rather than sfermions).

From Eq.~\eqref{gauginomass} we can read off the gaugino mass in the metastable vacuum (at $X=0$) to be
\begin{equation}
\label{name}
\bar{m}_{\lambda}=\Tr(\mu^{-1}\lambda)\zeta\leq N_{mess}\kappa_{max}\zeta,
\end{equation}
where $\kappa_{max}$ is the largest eigenvalue of the matrix $\mu^{-1}\lambda$.

The mass matrix for the fermionic messengers is,
\begin{equation}
(\lambda_{ab} X+\mu_{ab})=\mu_{ac}((\mu^{-1}\lambda)_{cb}X+\delta_{cb}).
\end{equation}
From this one can directly read off that there is a massless fermion at
\begin{equation}
X_{\star}=-\frac{1}{\kappa_{max}}.
\end{equation}
As argued in {\bf \S3}, one of the bosonic messengers is tachyonic there.
Therefore the minimal distance $\Delta X$ to a state of lower energy is less than $|X_{\star}|$.
Taking into account the (classical) flatness of the potential in the $X$-direction an estimate for the tunneling rate is
\begin{equation}
{\rm rate}\sim \exp\left(-const \cdot\left(\frac{\Delta X}{\sqrt{\zeta}}\right)^4\right),
\end{equation}
where $\Delta X$ is compared to the scale of the vacuum energy, given by $\sqrt{\zeta}$. We have
\begin{equation}
\label{stability}
\frac{|\Delta X|}{\sqrt{\zeta}}\leq \frac{|X_{\star|}}{\sqrt{\zeta}}=\frac{1}{\sqrt{\zeta}|\kappa_{max}|}
\leq N_{mess}\frac{\sqrt{\zeta}}{\bar{m}_{\lambda}}=N_{mess} \frac{\sqrt{m_{3/2}M_{P}}}{\bar{m}_{\lambda}}\sim \frac{g^2}{16\pi^2}\frac{\sqrt{m_{3/2}M_{P}}}{m_{\lambda}}.
\end{equation}
Here, we have used Eq.~\eqref{name} and the relation between the underlying SUSY breaking scale and the gravitino mass, $m_{3/2}=\zeta/M_{P}$. In the last
step we have included the factor $g^2/(16\pi^2)$ relating the gaugino mass scale to the actual gaugino masses.

Equation \eqref{stability} directly relates the distance between the vacua to the measurable gaugino and gravitino masses.
The longevity of the metastable vacuum
requires, \linebreak $\Delta X/\sqrt{\zeta}\gtrsim {\mathcal O}(1)$.
Using the lower bound $m_{\lambda}\gtrsim 100\,{\rm GeV}$ therefore requires a minimal gravitino mass of $\sim 10\,{\rm eV}$.
This nicely complements constraints from cosmology. Future measurements of the gaugino and gravitino masses (assuming their existence)
could shed light on the lifetime of the vacuum.

\section*{Acknowledgements}
We thank Matthew Dolan for interesting discussions.


\newpage

\begin{thebibliography}{10}
\bibitem{Dine:1981gu}
  M.~Dine and W.~Fischler,
  ``A Phenomenological Model Of Particle Physics Based On Supersymmetry,''
  Phys.\ Lett.\  B {\bf 110} (1982) 227.

\bibitem{Dine:1982zb}
  M.~Dine and W.~Fischler,
  ``A Supersymmetric Gut,''
  Nucl.\ Phys.\  B {\bf 204} (1982) 346.

\bibitem{AlvarezGaume:1981wy}
  L.~Alvarez-Gaume, M.~Claudson and M.~B.~Wise,
  ``Low-Energy Supersymmetry,''
  Nucl.\ Phys.\  B {\bf 207} (1982) 96.

\bibitem{Nappi:1982hm}
  C.~R.~Nappi and B.~A.~Ovrut,
  ``Supersymmetric Extension Of The SU(3) X SU(2) X U(1) Model,''
  Phys.\ Lett.\  B {\bf 113} (1982) 175.


\bibitem{ISS}
K.~Intriligator, N.~Seiberg and D.~Shih,
``Dynamical SUSY breaking in meta-stable vacua,''
JHEP {\bf 0604} (2006) 021   [hep-th/0602239].

\bibitem{Giudice:1998bp}
  G.~F.~Giudice and R.~Rattazzi,
  ``Theories with gauge-mediated supersymmetry breaking,''
  Phys.\ Rept.\  {\bf 322}, 419 (1999)
  [arXiv:hep-ph/9801271].



\bibitem{Murayama:2006yf}
  H.~Murayama and Y.~Nomura,
  ``Gauge mediation simplified,''
  Phys.\ Rev.\ Lett.\  {\bf 98} (2007) 151803
  [arXiv:hep-ph/0612186].

\bibitem{Izawa:1997gs}
  K.~I.~Izawa, Y.~Nomura, K.~Tobe and T.~Yanagida,
  ``Direct-transmission models of dynamical supersymmetry breaking,''
  Phys.\ Rev.\  D {\bf 56} (1997) 2886
  [arXiv:hep-ph/9705228].

\bibitem{Kitano:2006xg}
  R.~Kitano, H.~Ooguri and Y.~Ookouchi,
  ``Direct mediation of meta-stable supersymmetry breaking,''
  Phys.\ Rev.\  D {\bf 75} (2007) 045022
  [arXiv:hep-ph/0612139].

\bibitem{Csaki:2006wi}
  C.~Csaki, Y.~Shirman and J.~Terning,
  ``A simple model of low-scale direct gauge mediation,''
  JHEP {\bf 0705} (2007) 099
  [arXiv:hep-ph/0612241].

\bibitem{Abel:2007jx}
S.~Abel, C.~Durnford, J.~Jaeckel and V.~V.~Khoze,
``Dynamical breaking of $U(1)_{R}$ and supersymmetry in a metastable vacuum,''
Phys.\ Lett.\  B {\bf 661} (2008) 201   [arXiv:0707.2958 [hep-ph]].


\bibitem{Abel:2007nr}
S.~A.~Abel, C.~Durnford, J.~Jaeckel and V.~V.~Khoze,
``Patterns of Gauge Mediation in Metastable SUSY Breaking,''
JHEP {\bf 0802} (2008) 074   [arXiv:0712.1812 [hep-ph]].

\bibitem{Abel:2008gv}
  S.~Abel, J.~Jaeckel, V.~V.~Khoze and L.~Matos,
  ``On the Diversity of Gauge Mediation: Footprints of Dynamical SUSY
  Breaking,''
  JHEP {\bf 0903}, 017 (2009)
  [arXiv:0812.3119 [hep-ph]].

\bibitem{Komargodski:2009jf}
  Z.~Komargodski and D.~Shih,
  ``Notes on SUSY and R-Symmetry Breaking in Wess-Zumino Models,''
  JHEP {\bf 0904} (2009) 093
  [arXiv:0902.0030 [hep-th]].

\bibitem{Shihextraordinary}
  C.~Cheung, A.~L.~Fitzpatrick and D.~Shih,
  ``(Extra)Ordinary Gauge Mediation,''
  JHEP {\bf 0807} (2008) 054
  [arXiv:0710.3585 [hep-ph]].

\bibitem{Meade:2008wd}
  P.~Meade, N.~Seiberg and D.~Shih,
  ``General Gauge Mediation,''
  arXiv:0801.3278 [hep-ph].

\bibitem{Carpenter:2008wi}
L.~M.~Carpenter, M.~Dine, G.~Festuccia and J.~D.~Mason,
``Implementing General Gauge Mediation,''   arXiv:0805.2944 [hep-ph].

\bibitem{Seiberg:2008qj}
  N.~Seiberg, T.~Volansky and B.~Wecht,
  ``Semi-direct Gauge Mediation,''
  JHEP {\bf 0811} (2008) 004
  [arXiv:0809.4437 [hep-ph]].



\bibitem{Buican:2008ws}
  M.~Buican, P.~Meade, N.~Seiberg and D.~Shih,
  ``Exploring General Gauge Mediation,''
  JHEP {\bf 0903} (2009) 016
  [arXiv:0812.3668 [hep-ph]].


\bibitem{Carpenter:2008he}
  L.~M.~Carpenter,
  ``Surveying the Phenomenology of General Gauge Mediation,''
  arXiv:0812.2051 [hep-ph].

\bibitem{Rajaraman:2009ga}
  A.~Rajaraman, Y.~Shirman, J.~Smidt and F.~Yu,
  ``Parameter Space of General Gauge Mediation,''
  arXiv:0903.0668 [hep-ph].


\bibitem{Giveon:2009yu}
  A.~Giveon, A.~Katz and Z.~Komargodski,
  ``Uplifted Metastable Vacua and Gauge Mediation in SQCD,''
  arXiv:0905.3387 [hep-th].

\bibitem{Abel:2006cr}
  S.~A.~Abel, C.~S.~Chu, J.~Jaeckel and V.~V.~Khoze,
  ``SUSY breaking by a metastable ground state: Why the early universe
  preferred the non-supersymmetric vacuum,''
  JHEP {\bf 0701} (2007) 089
  [arXiv:hep-th/0610334].

\bibitem{Craig:2006kx}
  N.~J.~Craig, P.~J.~Fox and J.~G.~Wacker,
  ``Reheating metastable O'Raifeartaigh models,''
  Phys.\ Rev.\  D {\bf 75} (2007) 085006
  [arXiv:hep-th/0611006].


\bibitem{Fischler:2006xh}
  W.~Fischler, V.~Kaplunovsky, C.~Krishnan, L.~Mannelli and M.~A.~C.~Torres,
  ``Meta-Stable Supersymmetry Breaking in a Cooling Universe,''
  JHEP {\bf 0703} (2007) 107
  [arXiv:hep-th/0611018].


\bibitem{Abel:2006my}
  S.~A.~Abel, J.~Jaeckel and V.~V.~Khoze,
  ``Why the early universe preferred the non-supersymmetric vacuum. II,''
  JHEP {\bf 0701} (2007) 015
  [arXiv:hep-th/0611130].

\bibitem{Nelson:1993nf} A.~E.~Nelson and N.~Seiberg,
``R symmetry breaking versus supersymmetry breaking,''
 Nucl.\ Phys.\  B {\bf 416}, 46 (1994)   [arXiv:hep-ph/9309299].


\bibitem{Abel:2008tx}
  S.~Abel and V.~V.~Khoze,
  ``Direct Mediation, Duality and Unification,''
  JHEP {\bf 0811}, 024 (2008)
  [arXiv:0809.5262 [hep-ph]].


\bibitem{Cho:2007yn}
H.~Y.~Cho and J.~C.~Park,
``Dynamical $U(1)_R$ Breaking in the Metastable Vacua,''
 JHEP {\bf 0709} (2007) 122   [arXiv:0707.0716 [hep-ph]].



\bibitem{Sun:2008va}
  Z.~Sun,
  ``Tree level Spontaneous R-symmetry breaking in O'Raifeartaigh models,''
  JHEP {\bf 0901} (2009) 002
  [arXiv:0810.0477 [hep-th]].


\bibitem{Ray:2006wk}
  S.~Ray,
  ``Some properties of meta-stable supersymmetry-breaking vacua in Wess-Zumino
  models,''
  Phys.\ Lett.\  B {\bf 642} (2006) 137
  [arXiv:hep-th/0607172].




\end{thebibliography}
\end{document}